\documentstyle[a4,12pt]{article}
\begin{document}
\titlepage
\begin{flushright}
September 1997\\
LPTM 97/44\\
hep-th/yymmxxx
\end{flushright}
\vskip 1cm
\begin{center}
{\Huge \bf Anomalies of the SO(32) five-brane
and their cancellation}
\end{center}
\vskip 2cm
\begin{center}
{ \large{J. Mourad}\footnote{e-mail: mourad@qcd.th.u-psud.fr}}
\end{center}

\begin{center}
Laboratoire de Physique Th\'eorique et Mod\'elisation\\
Universit\'e de Cergy-Pontoise, 
Site Saint-Martin\\ F-95302 Cergy-Pontoise, France
\end{center}
\vskip 2cm
\begin{center}
{\large \bf Abstract}
\end{center}
The anomaly due to the chiral fermions on the world-volume of the $SO(32)$ five-brane
is calculated. 
It is shown that this contribution has the correct structure for it to be cancelled by the variation of the classical world-volume action. 
The cancellation mechanism requires a Green-Schwarz-like
term in the classical action. The result confirms the field content of the $SO(32)$ five-brane proposed by Witten.    
\newpage
\section{Introduction}
Anomaly cancellation in ten dimensions
greatly restricts the number of consistent 
string theories. 
The Green-Schwarz mechanism \cite{g} allows 
the anomaly cancellation for the $SO(32)$
and $E_8\times E_8$ string theories.
The essential ingredient in the Green-Schwarz mechanism is that the anomaly polynomial
of the effective target space action
be factorizable as $I_{12}=X_4 X_8$.
The $X_4$ part enters then in the modification of the Bianchi identity of the three-form field strength: $dH=X_4$
and the $X_8$ part in the modification of the equations of motion $d^{*}H=X_8$
\footnote{The analysis in  the Introduction is sketchy,  the dependance on the various constants and the dilaton will be explicited in section 4}.
These two modifications ensure that the total effective action is anomaly free.
Very interestingly, there exists a way to understand the modification of the Bianchi identity from the heterotic string world-sheet point of view \cite{h}. In fact the pull-back of $X_4$
into the world-sheet is simply the anomaly polynomial of the two-dimensional sigma model. The cancellation of the world-sheet anomalies requires a non-trivial variation 
of the two-form which  gives rise to the modified Bianchi identity and which via the interaction term
\begin{equation}
\int_{\Sigma_2} B,
 \end{equation}
compensates the anomaly.  

A way to understand the modification of the 
equations of motion of the three-form 
was proposed \cite{du1,di,i,b,l} and relies on the
existence in string theory of a solitonic extented object
of five spatial dimensions
that preserves half of the supersymmetries: the heterotic five-brane \cite{du2,s,c}. In fact the five-brane is expected to couple to the six-form, $B_6$,
by the world-volume term
\begin{equation}
\int_{\Sigma_6} B_6,
\end{equation}
where $B_6$ is the dual to $B_2$ that is
\begin{equation}
H_7={*}H,
\end{equation}
$H_7$ being  the field strength of $B_6$.
Since the five-brane is chiral,
world-volume anomalies of the five-brane
are expected to appear. Their cancellation
would require a non-trivial transformation of $B_6$ which in turn modifies the Bianchi identity of $H_7$ that is the equations of motion of $H$. So the $X_8$ part is expected to be related to the world-volume anomaly of the five-brane.
The precise relation will be explained later on.
This program has not been fully accomplished \cite{l} because the world-volume
fermions of the heterotic five-brane were not all known. 

Recently, Witten \cite{w} proposed
a field content of the SO(32) five-brane.
This field content comprises an $SU(2)$
vector supermultiplet and a pseudo-real hypermultiplet belonging to the fundamental representations of $SU(2)$ and $SO(32)$.
The aim of this paper is to re-examine 
the anomaly issue of the five-brane in the light of this new field content of the $SO(32)$ five-brane. We shall prove that indeed the anomaly of the five-brane has the correct structure. This is due to a number of cancellations and ``miracles"
which have a tiny chance to happen
for a different field content of the five-brane. The result constitutes thus a confirmation of Witten's proposition. 
 
In Section 2 we briefly review the field content of the $SO(32)$ five-brane 
proposed by Witten.
In Section 3 we calculate the contribution
of the various world-volume chiral fermions 
to the anomaly. The local symmetries 
that are potentially anomalous are 
$i)$ those of a bare six-dimensional theory: Lorentz $SO(1,5)$ rotations  and $SU(2)$ gauge transformations and
$ii)$ those that are induced 
from ten-dimensional symmetries: the $SO(4)$ rotations of the normal bundle and $SO(32)$ gauge transformations.   
We shall express the anomaly eight-form in terms of characteristic classes of the four bundles 
involved. We  find terms that depend on the Euler class of the normal bundle not encoutered before. 
In Section 4
we examine the anomaly cancellation mechanism. Since the $SO(32)$ five-brane has no antisymmetric tensor field, no
purely six-dimensional Green-Schwarz-like term is possible
in the effective world-volume action. However it is possible
to have  a term that depends on the restriction to the five-brane 
of the ten-dimensional antisymmetric 
tensor field. 
This renders the anomaly cancellation more constraining  than
is the case for a six-dimensional theory 
with a tensor field \cite{e,sh}.
We shall find that the anomaly calculated in section 3 is indeed of this constrained form. In section 5 we briefly consider the generalisation to the case of $k$ coincident five-brane and finally in section 6 we collect our conclusions.

\section{The $SO(32)$ five-brane}

According to the heterotic-type I duality,
the heterotic $SO(32)$ five-brane  
is mapped to the Dirichlet five-brane of the type I theory \cite{wp}.
 D-branes present the advantage of having an explicit conformal theory description \cite{p}.
In fact, the excitations of the
D-branes are open strings that have
one boundary or two boundaries that end on the D-brane. All that one have to know is the Chan-Paton factors characterizing the 
open strings that end on the D-five-brane.
Motivated by the  physics of the instanton  
of zero size, Witten \cite{w} proposed that the 
Chan-Paton factors are those of $Sp(1)=SU(2)$.
This was confirmed later by Gimon and Polchinsky \cite{gp}. 

The determination of the spectrum of the D-five-brane goes along the same line as the usual  open string
except that one has to consider two sectors
depending on whether there is one (DN sector) or both ends (DD sector) of the string on the D-brane. Furthermore, the operator $\Omega$ that exchanges the two ends of the string acts in an unusual manner. We refer the reader to \cite{w,wp} for the details. The result is that from the DD sector one gets: an $Sp(1)$ gauge field, $G$, 
its supersymmetric partner, $\lambda$,  which
belong to the $4_-$ of $SO(1,5)$ and the
$2_-$ of $SO(4)$, and finally four scalars
which represent the fluctuations of the five-brane in ten dimensions and their supersymmetric partners, $\theta$ which transform in
the $4_+$ of $SO(1,5)$ and the $2_+$ of 
$SO(4)$. Note the probably {\it a priori} unexpected transformation of the gauginos under the $SO(4)$ group. The 
transformation of $\theta$ under the $SO(4)$ group can be understood as
follows: the Green-Schwarz fermionic coordinates of the five-brane are a Weyl-Majorana fermion in ten dimensions whose reduction to six dimensions
transforms in the $(4_+,2_+)+(4_-,2_-)$
of $SO(1,5)\times SO(4)$, since half of the supersymmetries are broken
we are left with the
$(4_+,2_+)$ component. An important fact for the anomaly cancellation is  that $\lambda$ and $\theta$ have opposite chirality, this is due to $N=1$ six-dimensional supersymmetry.

Since the end of the string on the D-five-brane carries a $2$ of $SU(2)$ and the end with Neumann boundary conditions carries a $32$ of $SO(32)$ the states 
from the DN sector must be in the (32,2)
of $SO(32)\times SU(2)$. It turns out that  
from that sector one gets one pseudo-real hypermultiplet belonging to this representation. The scalars of this hypermultiplet transform in the vector representation of $SO(4)$ and the fermions, $\psi$,  transform under the $4_+$ of $SO(1,5)$ and are singlets under the $SO(4)$.

\section{The anomaly calculation}

In this section we shall calculate the contributions to the anomaly of the three fermion multiplets described in the preceding section. We shall use the standard anomaly formulas as found for example in \cite{a}.
The transformations of interest are: $SO(1,5)$ Lorentz 
transformations on $T(\Sigma)$, the tangent bundle of the five-brane,
$SO(4)$ Lorentz transformations 
on $N$, the normal bundle of the five-brane, 
$SO(32)$ gauge transformations and finally 
the $SU(2)$ gauge transformations. We will denote by
$2\pi F$ the  $SO(32)$ curvature  
and by $2\pi G$ the 
$SU(2)$ curvature. The anomaly in $D$ dimensions is given by a 
$(D+2)$-form: $\hat{I}_{D+2}=2\pi I_{D+2}$.
The $2\pi$ factors are
included to simplify the anomaly formulas.
We shall also express the anomaly in terms of characteristic classes of the 
ten dimensional tangent bundle $TQ$, and of the four dimensional normal bundle $N$.

The fermions, $\theta$, describing the fermionic coordinates of the five-brane
are symplectic-Weyl-Majorana spinors
that belong to the $(4_+,2_+)$ of $SO(1,5)\times SO(4)$.
The associated anomaly
reads:
\begin{equation}
I_8^\theta={{1}\over{2}}{\hat A}(T\Sigma)chS_+(N),
\end{equation}
where we keep only, as usual, terms of degree eight; $\hat A$ is the
Dirac
genus, $ch$ is the Chern character and $S_+(N)$
is the spin bundle with positive chirality.
The factor $1/2$ is present
because the fermions are symplectic-Majorana spinors.
The Dirac genus is given by
\begin{equation}
{\hat A}(T\Sigma)=1-{{p_1(\Sigma)}\over{24}}+
{{7p_1^2(\Sigma)-4p_2(\Sigma)}\over{5760}}+\dots,\label{ge}
\end{equation}
and the Chern character by
\begin{equation}
ch(S_{\pm}(N))=2+{{p_1(N)\pm 2\chi(N)}\over{4}}+
{{p_1^2(N)+4p_2(N)\pm 4p_1(N)\chi(N)}\over{192}}.\label{ch}
\end{equation}
In equations (\ref{ge}) and (\ref{ch}) $p_i$ denotes the $i^{th}$
Pontryagin class and $\chi$ the Euler class. They are related to the curvature $\Omega$ of $N$ by
\begin{eqnarray}
p_1(N)=-{{1}\over{2(2\pi)^2}}tr(\Omega^2),\quad p_2(N)={{1}\over{8(2\pi)^4}}
\left( (tr(\Omega^2))^2-2tr(\Omega^4)\right),
\nonumber\\ 
\chi(N)={{1}\over{8(2\pi)^2}}\epsilon_{abcd}\Omega^{ab}\wedge\Omega^{cd}.
\end{eqnarray}
It will we convenient to express the anomaly in terms of ten-dimensional
quantities and the curavature of the normal bundle. 
Let $TQ$ denotes the restriction
of the ten-dimensional tangent space to $\Sigma$ then
since $T\Sigma+N=TQ$, we have $p_1(\Sigma)=p_1(Q)-p_1(N)$
and $p_2(\Sigma)=p_2(Q)-p_2(N)-p_1(N)p_1(Q)+p_1^2(N)$.
Keeping  only terms of order eight we get
\begin{eqnarray}
I^\theta_8=  {{7p_1^2(Q)-4p_2(Q)}\over{5760}}-
{{2p_1(Q)p_1(N)+3p_1(Q)\chi(N)}\over{288}}\nonumber\\
+16{{4p_2(N)+3p_1^2(N)}
\over{5760}}+{{p_1(N)\chi(N)}\over{48}}.
\end{eqnarray}
The $SU(2)$ gauginos
are symplectic-Weyl-Majorana that belong to the 
$(1,3,4_-,2_-)$ of $SO(32)\times  SU(2)\times SO(1,5)\times SO(4)$. Their
contribution to the anomaly reads
\begin{equation}
I_8^\lambda=-{{1}\over{2}}{\hat A}(T\Sigma)ch S_-(N)Tr(e^{iG}),
\end{equation}
where $Tr$ is the trace in the adjoint representation of $SU(2)$ and the minus sign reflects the fact that $\lambda$ has negative chirality.
Keeping terms of order eight we obtain
\begin{eqnarray}
I_8^\lambda=
-3{{7p_1^2(Q)-4p_2(Q)}\over{5760}}
+3{{2p_1(Q)p_1(N)-3p_1(Q)\chi(N)}
\over{288}}\nonumber\\
-
{{p_1(Q)Tr(G^2)}\over{48}}
-48{{3p_1^2(N)+4p_2(N)}\over{5760}}+
3{{p_1(N)\chi(N)}\over{48}}\nonumber\\+
{{p_1(N)Tr(G^2)}\over{12}}-
{{\chi(N)Tr(G^2)}\over{8}}
-{{Tr (G^4)}\over{24}}.
\end{eqnarray}

The hypermultiplet contains 128 real scalars and
32 symplectic-Weyl-Majorana fermions transforming 
in the (32,2) of $SO(32)\times SU(2)$. They have the 
opposite chirality compared to the gaugino
and are singlets under the $SO(4)$ group. Their contribution to the anomaly  reads:
\begin{equation}
 I_8^{\psi}={{1}\over{2}}{\hat A}(T\Sigma)tr(e^{iF})tr(e^{iG}),\label{xx}
\end{equation}
where $tr$ is the trace in the fundamental representation,
 the factor $1/2$ is due to the reality condition.
The evaluation of (\ref{xx}) gives
\begin{eqnarray}
I_8^\psi=
32{{7p_1^2(Q)-4p_2(Q)}\over{5760}}
-16{{p_1(Q)p_1(N)}\over{288}}\nonumber\\
+
16{{p_1(Q)tr(G^2)}\over{48}}
+{{p_1(Q)tr(F^2)}\over{48}}
\nonumber\\
+32{{3p_1^2(N)+4p_2(N)}\over{5760}}-
16{{p_1(N)tr(G^2)}\over{48}}
-{{p_1(N)tr(F^2)}\over{48}}\nonumber\\
+
16{{tr (G^4)}\over{24}}+
{{tr(F^4)}\over{24}}+
{{tr(G^2)tr(F^2)}\over{8}}.
\end{eqnarray}
 The traces in the fundamental 
representations and in tha adjoint representations of $SU(2)$ are related by
\begin{equation}
Tr(G^2)=4 tr(G^2),\quad
Tr(G^4)= 16 tr(G^4).\label{rel}
\end{equation}
Summing the contributions of $\theta$, $\lambda$ and $\psi$ and using the relations (\ref{rel}) we get for the total anomaly
\begin{eqnarray}
I_8=I_8^\theta+I_8^\lambda+I_8^\psi= {{7p_1^2(Q)-4p_2(Q)}\over{192}}
-{{p_1(Q)p_1(N)}\over{24}}
-{{p_1(Q)\chi(N)}\over{24}}\nonumber\\
+{{p_1(Q)tr(G^2)}\over{4}}+
{{p_1(Q)tr(F^2)}\over{48}}
+{{p_1(N)\chi(N)}\over{12}}
-{{p_1(N)tr(F^2)}\over{48}}\nonumber\\
-{{\chi(N)tr(G^2)}\over{2}}
+{{tr(F^4)}\over{24}}+
{{tr(G^2)tr(F^2)}\over{8}}.\label{tot}
\end{eqnarray}
Note that the terms in $tr(G^4),
\ p_1(N)tr(G^2),\ p_2(N)$
 and $p_1^2(N)$  cancelled.
This allows the partial factorization
\begin{eqnarray}
I_8=\left[\chi(N)-{{tr(F^2)}\over{4}}
-{{p_1(Q)}\over{2}}\right]
\left[{{p_1(N)}\over{12}}-{{p_1(Q)}\over{24}}-{{tr(G^2)}\over{2}}\right]\nonumber\\
+{{3p_1^2(Q)-4p_2(Q)}\over{192}}+
{{p_1(Q)tr(F^2)}\over{96}}+{{tr(F^4)}\over{24}}.\label{fa}
\end{eqnarray}
We shall see in the next section that this partial factorization is of the required form for the anomalies to cancel.

\section{Anomaly cancellation}
Since anomaly cancellation in ten dimensions requires a non-trivial transformation of $B$ and $B_6$ and since the world-volume action of the five-brane 
depends on these forms, the classical action of the five-brane is not  
expected to be  invariant. So classical anomalies of the world-volume action may compensate the quantum anomalies computed in the preceding section. Here, we shall
first determine, from the known ten-dimensional 
transformations of $B$ and $B_6$, the structure of the expected classical anomaly and then prove that total anomaly cancellation is possible.
 
Recall that the ten dimensional space-time
anomalies are given by a twelve form,
$I_{12}$, which factorizes as
\begin{equation}
I_{12}=X_4X_8,\label{an}
\end{equation}
with
\begin{equation}
X_4=-{{tr(F^2)}\over{4}}
-{{p_1(Q)}\over{2}},
\end{equation}
and
\begin{equation}
X_8={{3p_1^2(Q)-4p_2(Q)}\over{192}}+
{{p_1(Q)tr(F^2)}\over{96}}+
{{tr(F^2)}\over{24}}.
\end{equation}
Anomaly cancellation in ten dimensions requires that the Bianchi identity of the three form field strenght be modified
to read
\begin{equation}
dH=\alpha'X_4,
\end{equation}
where $\alpha'$ is the string slope,
and that the action be supplemented with the Green-Schwarz term
\begin{equation}
S_{GS}=T_2\int_{Q}\ B \wedge X_8,\label{gs}
\end{equation}
where $T_2$ is the string tension: $T_2\alpha'=2\pi$.
The modification of the Bianchi identity implies that the two-form transforms as
\begin{equation}
\delta B=-\alpha'X_2^1,
\end{equation}
where $X_2^1$ is related to $X_4$ by the descent equations
\begin{equation}
X_4=dX_3,\ \ \delta X_3=dX_2^1.
\end{equation} 
It is easy then to see that the variation of the 
term (\ref{gs}) compensates the anomaly (\ref{an}).
Due to the Green-Schwarz term,
 the modified equations of motion of the two-form
become
\begin{equation}
d (e^{-\phi\ *}H)=2T_2\kappa^2X_8,
\end{equation}
where $\phi$ is the dilaton, $^{*}$ denotes the Hodge dual in the string metric and
$\kappa$ is the ten-dimensional gravitational constant.

The five-brane couples to the dual,
$B_6$, of the two-form $B$. This coupling is realised by the term
\begin{equation}
T_6\int_{\Sigma}B_6, \label{int}
\end{equation}
$T_6$ being the five-brane tension.
The forms $B_6$ and $B$
are related, via their field strength, by
$H_7=e^{-\phi\ *}H$, which implies that
\begin{equation}
dH_7=2T_2\kappa^2X_8.
\end{equation}
The Bianchi identity of $H_7$ gives the transformation of   
$B_6$  as
\begin{equation}
\delta B_6=-2T_2\kappa^2 X_6^1.\label{vari}
\end{equation}
On the other hand  the Dirac quantization
condition  gives:
\begin{equation}
T_6T_2\kappa^2=n\pi,\label{di}
\end{equation} 
where $n$ is an arbitrary integer.
From equations (\ref{vari}) and (\ref{di}) we deduce that 
the interaction term (\ref{int}) varies as
\begin{equation}
-2\pi n\int_\Sigma \ X_6^1.\label{con1}
\end{equation}

Another term which may exist in the world-volume action is a Green-Schwarz-like term
\begin{equation}
\Delta S = T_2\int_\Sigma B\wedge Y_4,\label{gst}
\end{equation}
where $Y_4$ is a polynomial of degree four.
In the presence of the five-brane the Bianchi identity
of the three form $H$ is mofified and reads
\begin{equation}
dH=X_4+2T_6\kappa^{2}\delta(\Sigma),
\end{equation}
where $\delta(\Sigma)$ is a four-form defined by
\begin{equation}
\int_\Sigma\ \omega=\int_Q\ \omega\wedge\delta(\Sigma),
\end{equation}
for all six-forms $\omega$.
An important property noticed in \cite{w2} 
is that the restriction of $\delta(\Sigma)$
to $\Sigma$ has a finite part given by the Euler class of the normal bundle, that is
\begin{equation}
\delta(\Sigma)|_\Sigma=\chi(N).
\end{equation}
The variation of the restriction to $\Sigma$ of the two-form is then easily seen to be given by
\begin{equation}
\delta B|_\Sigma=-\alpha'X_2^1|_\Sigma-2T_6\kappa^{2}\chi_2^1(N).
\end{equation}
The corresponding variation of the Green-Schwarz-like
 term (\ref{gst}) is then
\begin{equation}
\delta \Delta S = -2\pi\int_\Sigma \Big(X_2^1+n\chi_2^1(N)\Big)\wedge Y_4.\label{con2}
\end{equation}
The two terms (\ref{con1}) and (\ref{con2}) contribute to the anomaly 
polynomial with 
\begin{equation}
-nX_8-\Big(X_4+n\chi(N)\Big)\wedge Y_4.
\end{equation}
We deduce that the total variation of the effective action vanishes if the 
quantum anomaly eight form $I_8$ is of the form
\begin{equation}
I_8=nX_8+\Big(X_4+n\chi(N)\Big)\wedge Y_4,\label{fo}
\end{equation}
for some integer $n$ and some polynomial $Y_4$.
This is the condition of anomaly cancellation on the five-brane.
Note that anomaly cancellation in a six-dimensional theory with an antisymmetric six-dimensional tensor field
\cite{e,sh} requires the anomaly to be factorizable as ${\tilde Y}_4\wedge Y_4$. The form (\ref{fo}) is much more constraining because it involves only one polynomial.

The partial factorization of the anomaly given in (\ref{fa}) is precisely of the form (\ref{fo}). The integer $n$ must be set to unity and  the polynomial  
$Y_4$ is given by
\begin{equation}
Y_4=-{{p_1(Q)}\over{24}}+
{{p_1(N)}\over{12}}-{{tr(G^2)}\over{2}}.\label{y}
\end{equation}
The effective world-volume action of the five-brane must thus contain the Green-Schwarz-like term (\ref{gst}) with $Y_4$ given by (\ref{y}). 
 
The fact that $I_8$ is of the correct form is due to a number of remarkable cancellation which seem miraculous.
In fact, the anomaly $I_8$ is {\it a priori} a sum of more than ten terms, requiring that it be of the form (\ref{fo})
with $Y_4$ arbitrary leads to a ``fine tuning" of more than five coefficients.

\section{$k$ coincident $SO(32)$ five-branes}
In this section we briefly consider the generalisation to the case of $k$
coincident five-branes. In this case, the Chan-Paton of the open strings ending on the D-five-branes  are those
of $Sp(k)$. The $Sp(1)$ gauge group of the single five-brane is replaced by $Sp(k)$.
The six dimensional zero modes living on the five-brane transform in the same representations as before
under $SO(1,5)\times SO(4)\times SO(32)$. 
The transformations under the gauge group $Sp(k)$ of the fermionic modes 
are as follows: $\theta$ is in the 
antisymmetric tensor product of two fundamental representations
$(k(2k-1)-1)+1$; $\lambda$ in the adjoint representation 
of dimension $k(2k+1)$ and finally $\psi$ is in the fundamental representation of dimension $2k$.

The contribution of each fermion multiplet can be calculated as in section 3.
We shall not write down the detailed contribution of each but give the final result of the total quantum anomaly:
\begin{eqnarray}
I_8=k{{7p_1^2(Q)-4p_2(Q)}\over{192}}
-k{{p_1(Q)p_1(N)}\over{24}}
-k^2{{p_1(Q)\chi(N)}\over{24}}\nonumber\\
+{{p_1(Q)tr(G^2)}\over{4}}+
k{{p_1(Q)tr(F^2)}\over{48}}
+k^2{{p_1(N)\chi(N)}\over{12}}
-k{{p_1(N)tr(F^2)}\over{48}}\nonumber\\
-k{{\chi(N)tr(G^2)}\over{2}}
+k{{tr(F^4)}\over{24}}+
{{tr(G^2)tr(F^2)}\over{8}}.\label{tota}
\end{eqnarray}
In obtaining this result, an essential use was made of the following trace identities
of $Sp(k)$:
\begin{eqnarray}
Tr(G^2)=(2k+2)tr(G^2),\quad Tr(G^4)=(2k+8)tr(G^4)+3(tr(G^2))^2,\nonumber\\
tr_{A}(G^2)=(2k-2)tr(G^2),\quad
tr_{A}(G^4)=(2k-8)tr(G^4)+3(tr(G^2))^2,
\end{eqnarray}
where $tr$, $Tr$ and $tr_{A}$ are respectively the trace in the fundamental
(2k), adjoint (k(2k+1))
and antisymmetric (k(2k-1)-1)
representations of $Sp(k)$.
Similarly to the case of a single five-brane, several terms cancelled
and the anomaly (\ref{tota}) can be written in the partially factorized form:
\begin{eqnarray}
I_8=\left[k\chi(N)-{{tr(F^2)}\over{4}}
-{{p_1(Q)}\over{2}}\right]
\left[k{{p_1(N)}\over{12}}-k{{p_1(Q)}\over{24}}-{{tr(G^2)}\over{2}}\right]\nonumber\\
+k{{3p_1^2(Q)-4p_2(Q)}\over{192}}+
k{{p_1(Q)tr(F^2)}\over{96}}+k{{tr(F^4)}\over{24}}.\label{fac}
\end{eqnarray}
The anomaly is again of the required form 
(\ref{fo}) but this time with $n=k$ and
\begin{equation}
Y_4=-k{{p_1(Q)}\over{24}}+
k{{p_1(N)}\over{12}}-{{tr(G^2)}\over{2}}.
\end{equation}
As expected, the tension of $k$ five-branes must be $k$ times the tension of a single five-brane.

\section{Conclusion}
We have proved that Witten's $SO(32)$ five-brane is anomaly free, this is an essential requirement for the consistency of the $SO(32)$ five-brane theory. As mentionned in the introduction, one can also interpret the result as a ``derivation" from the world-volume point of view of the 
Green-Schwarz mechanism. In fact, we can turn the argument around and from the requirement of world-volume anomaly cancellation get the transformations of $B$ and $B_6$ which imply the modified Bianchi identity and equations of motion of $H$.

In addition we have a prediction concerning the term (\ref{gst}) which must appear in the effective world-volume action of the five-brane. It would be interesting to have a confirmation of this result from the conformal theory point of view.
We have also proved that the system with $k$ coincident $SO(32)$ 
five-branes is also anomaly free. 
The field content of the $E_8\times E_8$ five-brane is still unknown and it is not clear whether there exists an  effective local field theory describing it. Anomaly cancellation arguments of the type 
exposed here may help the clarification
of this issue.

\end{document}